\documentclass[twocolumn,letter,traditabstract]{aa}
\usepackage{natbib}
\newcommand{\bibfont}{\tiny}
\usepackage{graphicx}
\usepackage[T1]{fontenc}
\usepackage{amsmath}
\usepackage{amssymb}
\usepackage{amsmath}
\usepackage{graphicx}
\usepackage{amsmath}
\usepackage{amssymb}
\usepackage{amsfonts}
\usepackage{graphicx,epsfig,graphics}
\usepackage{aas_macros}
\usepackage{graphics,hyperref,url}
\usepackage{color}
\usepackage{epstopdf}

\def\be{\begin{eqnarray}}
\def\ee{\end{eqnarray}}
\graphicspath{{}}

\begin{document}

\title{On preferred axes in WMAP cosmic microwave background data after subtraction of the integrated Sachs-Wolfe effect} 
\titlerunning{Preferred Axes in WMAP CMB Data}\author{A. Rassat \inst{1,2} \thanks{anais.rassat@epfl.ch} \and J.-L. Starck \inst{2}}
\institute{$^1$ Laboratoire d'Astrophysique, Ecole Polytechnique F\'ed\'erale de Lausanne (EPFL), Observatoire de Sauverny, CH-1290, Versoix, Switzerland.\\
$^2$ Laboratoire AIM, UMR CEA-CNRS-Paris, Irfu, SAp, CEA Saclay, F-91191 GIF-SUR-YVETTE CEDEX, France.}

\abstract
{There is currently a debate over the existence of claimed statistical anomalies in the cosmic microwave background (CMB), recently confirmed in \emph{Planck} data. Recent work has focussed on methods for measuring statistical significance, on masks and on secondary anisotropies as potential causes of the anomalies. We investigate simultaneously the method for accounting for masked regions and the foreground integrated Sachs-Wolfe (ISW) signal. We search for trends in different years of WMAP CMB data with different mask treatments. We reconstruct the ISW field due to the 2 Micron All-Sky Survey (2MASS) and the NRAO VLA Sky Survey (NVSS) up to $\ell=5$,   and we focus on the Axis of Evil (AoE) statistic and even/odd mirror parity, both of which search for preferred axes in the Universe. We find that removing the ISW reduces the significance of these anomalies in WMAP data, though this does not exclude the possibility of exotic physics. In the spirit of reproducible research, all reconstructed maps and codes will be made available for download at \url{http://www.cosmostat.org/anomaliesCMB.html}.}

\maketitle
\section{Introduction}
In recent years, several violations of statistical isotropy have been reported on the largest scales of the cosmic microwave background (CMB). A low quadrupole was reported in COBE data \citep{Hinshaw:1996,Bond:1998} and confirmed later with WMAP data \citep{Spergel:2003cb,w9:lowquad}. The octopole presented an unusual planarity and its phase seemed correlated with that of the quadrupole \citep{Tegmark:2003ve,OctPlanarity,Slosar:2004s,Copi:2010}. Other anomalies include a north/south power asymmetry \citep{Eriksen:2003db,Bernui:2006}, an anomalous cold spot \citep{Vielva:2004,Cruz:2005,Cruz:2006sv}, alignments of other large-scale multipoles \citep{Schwarz:2004,Copi:2005ff}, the so-called `Axis of Evil' \citep[AoE,][]{Land:2005ad} and other violations of statistical isotropy \citep{Hajian:2003s,Land:2004bs}.  Recently, \emph{Planck} has confirmed that these large-scale anomalies are still present in the CMB \citep{Planck:products,Planck:isotropy}, ruling out any origin due to a systematic in the data.

These anomalies are interesting because they point towards a possible violation of the standard model of cosmology, which predicts statistical isotropy and Gaussian fluctuations in the CMB, and therefore offer a window into exotic early-universe physics \citep[e.g.,][]{anomalies:exotic3,anomalies:exotic,anomalies:exotic2,anomalies:exotic4}. However, there is still much debate over the possible causes of these anomalies. Since these effects are on very large scales where there is large cosmic variance, the statistics used to measure the significance of the anomalies are subtle \citep{WmapAnomalies,Efstathiou:2010,ilc:w7,ilc:w9}. They could also be due to some foreground effects, which could be either contamination due to Galactic foreground effects \citep{ilc:w9,Kim:2012} or cosmological foregrounds that lead to secondary anisotropies in the CMB \citep{Rassat:2007KRL,Rudnick:2007,Hiranya:ksz,Smith:2010Huterer,CMBSN:2012,Francis:2010iswanomalies,Rassat:2012}.

\citet{Rassat:2012} have simultaneously investigated the impact of masks and the integrated Sachs-Wolfe (ISW) effect.  Missing data were accounted for with the sparse inpainting technique described in \cite*{Starck:2013}, which does not assume the underlying field is either Gaussian or isotropic, but allows for it to be. It was found that removing the ISW reduced the significance of two anomalies in WMAP data (the quadrupole/octopole alignment and the octopole planarity). In this work we focus on two anomalies, both related to preferred axes on the sky: the AoE effect and mirror parity. 
In Section \ref{sec:isw} we describe the reconstruction of the ISW field from 2MASS and NVSS data. In Section \ref{sec:results}, we search for violations of statistical isotropy before and after inpainting, as well as after both inpainting and ISW subtraction. In Section \ref{sec:discussion} we discuss our results and summarise our results compared with those from \citet{Rassat:2012}.

\section{Estimating the large-scale primordial CMB}\label{sec:isw}
\subsection{Theory}
Since statistical isotropy is predicted for the early Universe, analyses should focus on the primordial CMB, i.e. one from which secondary low-redshift cosmological signals have been removed. In Appendix \ref{app:isw}, we briefly review how to estimate the primordial CMB with a reconstructed map of the ISW effect. In practice, we estimate the primordial CMB on large scales by
\begin{equation} \hat{\delta}_{\rm prim}\simeq \delta_{\rm OBS} - \hat{\delta}^{\rm 2MASS}_{\rm ISW} - \hat{\delta}^{\rm NVSS}_{\rm ISW} - \delta_{kD,\ell=2},\label{eq:kd}\end{equation}
where $\hat{\delta}^{\rm 2MASS}_{\rm ISW}$ and $\hat{\delta}^{\rm NVSS}_{\rm ISW}$ are the estimated ISW contributions from the 2MASS and NVSS surveys \citep[see Section \ref{sec:data}, Appendix \ref{app:isw} and Section 2.2 from][for details on how these are estimated]{Rassat:2012}. The terms $\delta_{\rm OBS}$ and $\delta_{\rm prim}$ correspond to the observed and primordial CMB respectively. The term $\delta_{kD,\ell=2}$ is the temperature signal due to the kinetic Doppler quadrupole \citep{Copi:2005ff,Francis:2010iswanomalies}, for which we have produced publicly available maps in \citet{Rassat:2012}.

\subsection{Data}\label{sec:data}
We are interested in identifying trends and therefore considered WMAP renditions from various years, both before and after inpainting. These maps are described in Table 1 of \citet{Rassat:2012} (see also Appendix \ref{sec:cmb}).  The 2MASS \citep{Jarrett:2004a} and NVSS \citep{NVSS} data sets are described in Sections 3.2 and 3.3 or in \citet{Rassat:2012}.

We use the reconstructed ISW effect due to 2MASS and NVSS galaxies, using the method in \citet{Dupe:2011} and \citet{Rassat:2012}, but reconstructing the ISW fields up to $\ell=5$. The ISW reconstruction method is cosmology independent, and estimates the ISW amplitude directly from the CMB and galaxy field cross-correlations. This means that there is a different ISW map for each CMB map rendition considered. All map reconstructions are done using the publicly available {\tt ISAP} code, and the specific options used are described in detail in \citet{Rassat:2012}.

\begin{figure}[htbp]
   \centering
   	\vspace{-0.2cm}
            \includegraphics[width=9cm]{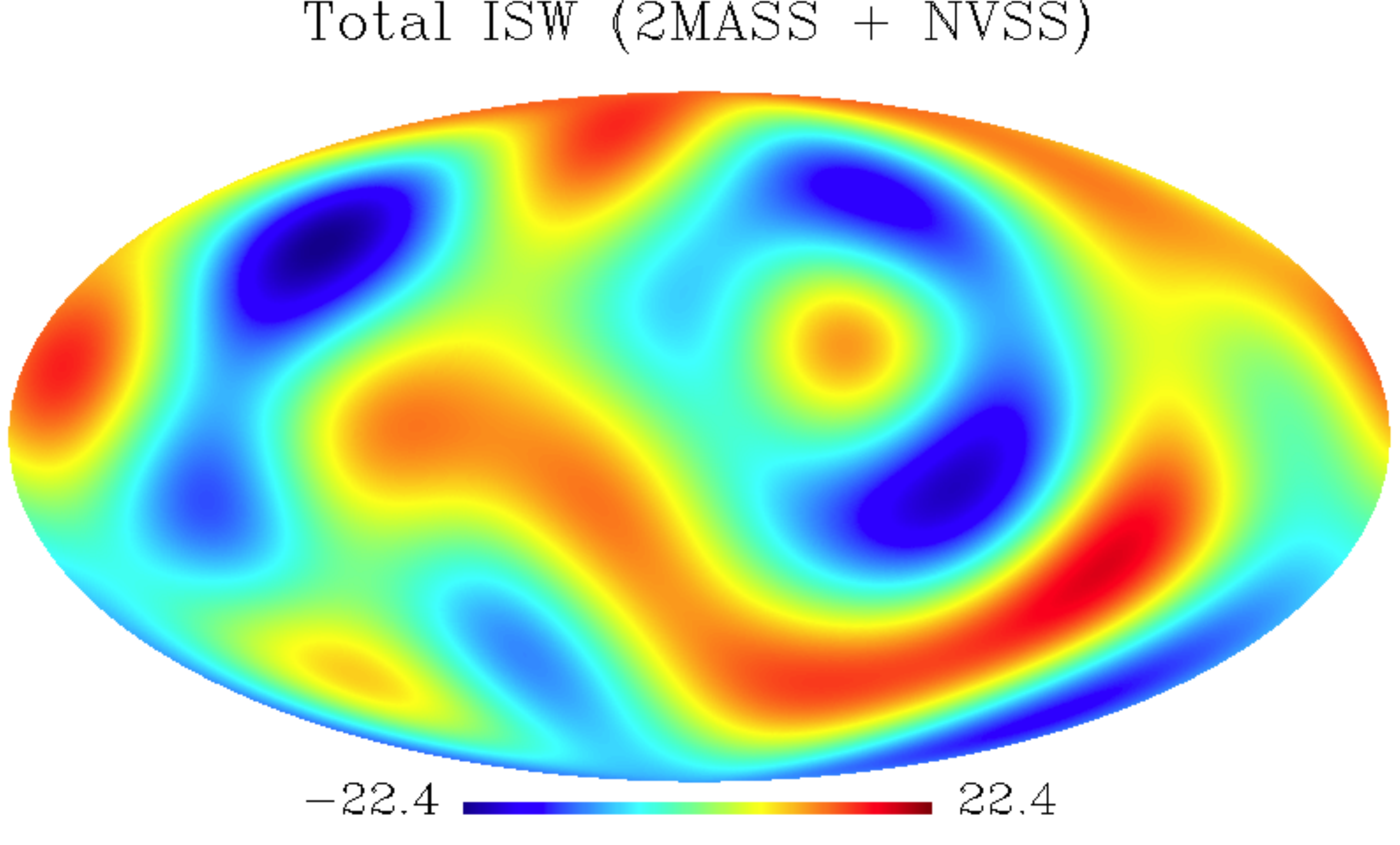} 
               	\vspace{-0.5cm}
   \caption{The large-scale ISW temperature field ($\ell=2-5$) due to 2MASS and NVSS galaxies. The amplitude is estimated directly from the data, including cross-correlation with WMAP9 data.}
   \label{fig:isw}
\end{figure}

The large-scale ISW temperature field ($\ell=2-5$) due to 2MASS and NVSS (where the amplitude is estimated from a correlation with WMAP9 data) is plotted in Figure \ref{fig:isw}. Since there is little redshift overlap between the surveys \citep[see Figure 3 in][]{Rassat:2012}, we estimate the ISW contribution from each survey independently and add the two resulting ISW maps to produce the map in Figure \ref{fig:isw}.

\begin{figure}[htbp]
   \centering
                  \vspace{-3.cm}
            \includegraphics[width=9cm]{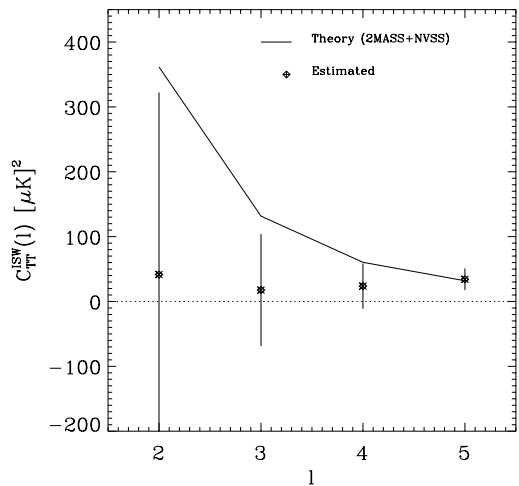} 
               \vspace{-3cm}
   \caption{Amplitude of the ISW temperature power $C_{\rm TT}^{\rm ISW}(\ell)$ (in $\mu K^2$) for $\ell=2-5$. The solid line is the theoretical prediction, and the data points are estimated from 2MASS and NVSS galaxies with WMAP9 data are shown with Gaussian error bars assuming $f_{\rm sky}=0.66$, i.e. the sky coverage of the NVSS survey.}
   \label{fig:ctt}
\end{figure}

In Figure \ref{fig:ctt}, we plot the amplitude of the ISW temperature power $C_{\rm TT}^{\rm ISW}(\ell)$ for $\ell=2-5$, which is measured directly from the map plotted in Figure \ref{fig:isw}, showing the theoretical prediction for a `Vanilla-model' cosmology and  taking contributions into account from both 2MASS and NVSS galaxies. Data points are shown for WMAP9, 2MASS, and NVSS data with Gaussian errors bars estimated analytically and assuming $f^{\rm NVSS}_{\rm sky}=0.66$ (i.e. the smallest $f_{\rm sky}$ value amongst the three maps). In general, we find that the ISW signal is below what is expected from theory for the fiducial cosmology we assumed.

\section{Preferred axes in WMAP data before and after inpainting and ISW subtraction}\label{sec:results}
We focus on both the AoE statistic and the even/odd mirror parity anomalies, which are described in detail in Appendix \ref{sec:theory}. In Appendix \ref{sec:validation}, we validate that sparse inpainting provides a bias-free reconstruction with which to study these anomalies using a large set of simulated CMB maps.

\subsection{Axis of Evil (AoE)}\label{sec:results:aoe}

\citet{Rassat:2007KRL} searched for an AoE directly in 2MASS data, hoping that this could link the measured anomaly to the ISW effect, but found no preferred axes in 2MASS data. Here, we can test an estimate of the primordial CMB estimated directly.

In Table \ref{tab:aoe} we report the preferred modes and axes for the 11 CMB maps considered in this paper. This table can be directly compared with Table 1 in \citet{Land:aoe2}, which reviewed the preferred modes and axes for W1 and W3 data. By analysing the `raw' maps (section 1 of Table \ref{tab:aoe}), we confirm their result that the preferred axis that was present in first-year data is not present in the ILC W3 data. We find the mean interangle is anomalously low for 4 WMAP renditions (TOH1, W5, W7, W9) with $\theta\sim 20^\circ$, but that for W3 data, $\theta\sim52^\circ$, similar to what \citet{Land:aoe2} found for W1 and W3 maps in their updated analysis of the AoE. As they found, the change occurs because the preferred mode for $\ell=2$ is $m=2$ for W1 (and also years 5 to 9), whereas for W3 data, the preferred mode for $\ell=2$ is $m=0$. They attribute this fact to the discontinuous nature of the AoE statistic, and underline how this feature constitutes a weakness in the AoE statistic.

However, following \citet{Francis:2010iswanomalies}, we subtract the kinetic Doppler (kD) from the public WMAP maps (i.e. by subtracting the last term in Equation \ref{eq:kd}) and re-perform the AoE analysis (section 2 of Table \ref{tab:aoe}, where only the quadrupole $\ell=2$ has changed). We find the statistic is now stable across all WMAP data considered, with the mean interangle spanning $\theta=18.5-20.8^\circ$.

The results after inpainting of the CMB maps (and kD subtraction) are presented in section 3 of Table \ref{tab:aoe}. We find inpainting has no effect on the preferred modes and directions for $\ell=2$ and $\ell=5$, which are generally unchanged and are the same for all maps, as was the case before inpainting. For $\ell=4$, the preferred mode and direction for all WMAP renditions now becomes similar to that for the TOH map before inpainting. The octopole ($\ell=3$) is the scale that changes the most after inpainting, both preferred mode and direction are changed for all WMAP renditions. The preferred modes and directions for $\ell=2-5$ are quite stable across all WMAP renditions after inpainting.  We note that when searching for quadrupole/octopole alignment, \citet{Rassat:2012} showed that the preferred axis of the octopole was stable after inpainting. However, this statistic enforced a search within a planar mode ($m=3$), which explains the difference in stability for the octopole. 

The change of only one preferred mode, notably the octopole in the W9 rendition that prefers the planar mode $m=3$ rather than $m=1$ for other WMAP renditions, has a significant effect on the mean interangle $\theta$ that changes from $48.3-51.0^\circ$ for W1-7 data to $16.3^\circ$, a value which is only found in $0.1\%$ of simulations. Further studies with improved (and smaller) masks, either on WMAP or \emph{Planck} data, could point to why the AoE statistic on W9 data is so different.

The results after inpainting and ISW subtraction are presented in section 4 of Table \ref{tab:aoe}. For the quadrupole, the preferred mode remains the same, and the preferred direction changes slightly. The scale with the most change are $\ell=3,4,5$, where both preferred mode and direction change for all WMAP renditions. In general, there is very good agreement across different WMAP renditions, except some differences for the octopole. After inpainting and ISW subtraction, the mean interangle now varies between $41.5^\circ-61.6^\circ$, which corresponds to 5.8\%-61.5$\%$ of values found in simulations. Even for the lowest value of $5.8\%$, the original anomalous alignment is no longer as significant as the previous value of $0.1\%$ on the ILC maps.

\subsection{Parity}
We calculate the $S_{\pm}$ values for each of the WMAP renditions using Equation \ref{eq:sodd}, and results are presented in Table \ref{tab:parity}, where the significance is calculated using 1000 full-sky Gaussian realisations. Before inpainting (part 1 of Table \ref{tab:parity}), we find no significant even ($S_+$) or odd ($S_-$) mirror parity, except for the TOH map, which shows a mild preference for even mirror parity (only 2.6\% of simulations have a higher $S_+$ value). After sparse inpainting (part 2 of Table \ref{tab:parity}), we find an increase in both even and odd mirror parity anomalies (except for the TOH map), with only $3.2-3.6\%$ of the simulations having a higher value for $S_-$ than both W7 and W9 maps and only $0.90-3.1\%$ of the simulations having a higher value for $S_+$. The significance of these remains at $<3\sigma$ though. We find the mirror-parity anomalies do not persist after sparse inpainting and subtraction of the reconstructed ISW signal (part 3 of Table \ref{tab:parity}), meaning the ISW could explain these anomalies.

\section{Discussion}\label{sec:discussion}

One of the main successes of the standard model of cosmology is its prediction of the Gaussian random fluctuations observed in the CMB. However, even since COBE data, several signatures of lack of statistic isotropy, or `anomalies', have been reported on large scales in WMAP data and recently confirmed in \emph{Planck} data \citep{Planck:products,Planck:isotropy}. Recent focus has been on testing the impact of different reconstruction methods or methods for dealing with Galactic foregrounds, while others have investigated how various foreground cosmological signals could affect these anomalies. 

In \citet{Rassat:2012}, we found that subtracting the ISW signal due to 2MASS and NVSS data from WMAP data lowered the significance of two previously reported anomalies: the quadrupole/octopole alignment and the octopole planarity.

\begin{table}[htbp]
   \centering
   \begin{tabular}{@{} lcc @{}} 
   \hline
Anomaly &After sparse&After ISW\\
&inpainting & subtraction\\
\hline
\emph{From \citet{Rassat:2012}}\\ 

Low quad&More anomalous & More anomalous\\

Quad/oct alignment&Less anomalous & Not anomalous\\
Oct planarity & Less anomalous & Not anomalous\\

\hline
\emph{This work}\\
Axis of evil&Less anomalous&Not anomalous\\
Even mirror parity &More anomalous&Not anomalous\\
Odd mirror parity &More anomalous&Not anomalous\\
\hline
   \end{tabular}
   \caption{Summary of results in this paper and \citet{Rassat:2012}. The anomalies in WMAP could be explained by the ISW effect, though other explanations remain possible.}
   \label{tab:summary}
   \vspace{-0.6cm}
\end{table}

In this work, we continued investigation of two other anomalies, both related to preferred axes in the sky: the Axis of Evil (AoE) and even/odd mirror parity in CMB data, i.e. parity with respect to reflections through a plane.  We first investigated whether sparse inpainting can be considered a bias-free reconstruction method for the two statistics and found that this is the case. We then applied sparse inpainting on various CMB maps up to $\ell=5$ and reconstructed the ISW maps from 2MASS and NVSS data also up to $\ell=5$. We considered the significance of the two reported anomalies, before inpainting, after inpainting, and both after inpainting and ISW subtraction. 

Our first approach to the AoE was to remove the kD quadrupole \citep[following][]{Francis:2010iswanomalies}, and we found that the AoE is consistent across all renditions of WMAP data (TOH, W3, W5, W7, and W9), unlike what \citet{Land:aoe2} found. After sparse inpainting, we found the AoE is no longer anomalous, mainly due to the change in preferred mode and axis of the octopole, except for WMAP9 data, where the anomaly persists. Further studies with improved (and smaller) masks, either on WMAP or \emph{Planck} data, could point to why the AoE statistic on WMAP9 data is so different. After sparse inpaiting, both even and odd mirror parities are increased in significance, but not enough to be considered significantly anomalous. 

We found that subtraction of the ISW effect due to 2MASS and NVSS galaxies, reduces the significance of these anomalies. These results, along with those in \citet{Rassat:2012} relating to the low quadrupole, the quadrupole/octopole alignment and the octopole planarity are summarised in Table \ref{tab:summary}. We note, however, that there are other signatures of statistical anomalies on large scales that we have not tested (e.g.: north/south asymmetry, cold spot, etc.) and that exotic physics remain possible.  These results are based on WMAP data alone, and should be repeated with \emph{Planck}.

\small{ In the spirit of reproducible research, all reconstructed maps and codes that constitute the main results of this paper will be made available for download at \url{http://www.cosmostat.org/anomaliesCMB.html}.}

\vspace{-0.2cm}
 \begin{acknowledgements}
The authors are grateful to the Euclid CMB cross-correlations working group for useful discussions. We use iCosmo\footnote{\url{http://www.icosmo.org}, \citet{Refregier:2011}}, Healpix \citep{healpix:2002,Gorski:2004by}, and ISAP\footnote{\url{http://jstarck.free.fr/isap.html}} software, along with 2MASS\footnote{\url{http://www.ipac.caltech.edu/2mass/}}, \emph{WMAP} \footnote{\url{http://map.gsfc.nasa.gov}}, and NVSS data\footnote{\url{http://heasarc.gsfc.nasa.gov/W3Browse/all/nvss.html}},
 and the Galaxy extinction maps of \citet{Schlegel:1997yv}.
 This research is in part supported by the European Research Council grant SparseAstro (ERC-228261) and by the Swiss National Science Foundation
(SNSF).
\end{acknowledgements}

\bibliographystyle{aa}
\bibliography{article.bib}

\vspace{-.4cm}
\appendix

\section{WMAP cosmic microwave background maps considered}\label{sec:cmb}
As we are interested in identifying trends in the data, we consider a suite of 11 different renditions of WMAP data:
the \citet{Tegmark:2003ve} reduced-foreground CMB Map (TOH1), the internal linear combination (ILC) WMAP maps from the 3rd year \citep[ILC W3,][]{ilc:w3}, 5th year \citep[ILC W5,][]{ilc:w5}, 7th year \citep[ILC W7,][]{ilc:w7}, and the 9th year \citep[ILC W9,][]{ilc:w9}, as well as sparsely inpainted versions of these maps. We also consider the sparsely inpainted WMAP ILC 5th year data reconstructed by \citet[Dela W5,][]{Delabrouille:2008} using wavelets. These are summarised in Table 1 of \citet{Rassat:2012}.

\section{Statistical anomalies and impact of sparse inpainting}\label{sec:theory}
\subsection{Preferred axis for low multipoles: Axis of Evil}\label{sec:theory:aoe}
It was first noted by \citet{OctPlanarity} that both the quadrupole and octopole of the CMB appeared planar (i.e. anomalously dominated by $m=\pm \ell$ modes) and were also aligned along a similar axis.

\citet{Land:2005ad} suggest searching for a more general axis by considering the power in \emph{any} mode $m$, instead of focussing on planar modes. This can be quantified by considering their statistic: \begin{equation} r_\ell =\max_{m, \hat{\bf n}} \frac{C_{\ell m}(\hat{\bf n})}{(2\ell+1)\hat{C}_\ell}.\label{eq:aoe}\end{equation}
The expressions $C_{\ell m}(\hat{\bf n})$ are given by $C_{\ell 0}(\hat{\bf n})=|a_{\ell 0}(\hat{\bf n})|^2$ and $C_{\ell m}(\hat{\bf n})=2|a_{\ell m}(\hat{\bf n})|^2$ for $m>0$ and $(2\ell+1)\hat{C}_\ell = \sum_m |a_{\ell m}|^2$, where $a_{\ell m}(\hat{\bf n})$ corresponds to the value of the $a_{\ell m}$ coefficients when the map is rotated to have $\hat{\bf n}$ as the $z$-axis. The above statistic finds both a preferred axis $\hat{\bf n}$ and a preferred mode $m$.

\citet{Land:2005ad} find that the preferred axes for $\ell=2,..., 5$ for WMAP 1 data seemed aligned along a similar axis in the direction of $(\ell, b) \sim (-100^\circ, 60^\circ)$, where the $l$ varied from $\simeq [-90^\circ,-160^\circ]$ and $b$ varied from $\simeq [48^\circ,62 ^\circ]$. By considering the mean interangle $\theta$ (i.e. the mean value of all possible angles between two axes $\hat{\bf n}_\ell$ and $\hat{\bf n}_{\ell'}$ for $\ell,\ell' = 2, ..., 5$ and $\ell \ne \ell'$), they find that only $0.1\%$ of simulations had a lower average value than the one measured in WMAP1 data ($\sim20^\circ$) and therefore rejected statistical isotropy at the 99.9\% confidence level. The preferred axis has been dubbed the ``AoE''.

\subsection{Mirror parity}
Another test is mirror parity in CMB data, i.e. parity with respect to reflections through a plane: $\hat{\bf x}=\hat{\bf x} - 2(\hat{\bf x}\cdot\hat{\bf n})\hat{\bf n}$, where $\hat{\bf n}$ is the normal vector to the plane. Since mirror parity is yet another statistic for which preferred axes can be found (i.e. the normal to the plane of reflection), it is complementary to the search for a preferred axis described in Section \ref{sec:theory:aoe}. 

In practice, mirror parity and preferred axes found using Equation \ref{eq:aoe} are statistically independent \citep{Land:2005odd}, so the coincidental presence of both increases the significance of the preferred axis anomaly. With all-sky data, one can estimate the $S$-map for a given multipole by\begin{equation} \tilde{S}_\ell(\hat{\bf n}) = \sum_{m=-\ell}^{\ell}(-1)^{\ell+m}\frac{|a_{\ell m}(\hat{\bf n})|^2}{{\hat C}_\ell}.\label{eq:paritymap}\end{equation}
Positive (negative) values of $\tilde{S}_\ell (\hat{\bf n})$ correspond to even (odd) mirror parities in the $\hat{\bf n}$ direction. 
The same statistic can also be considered summed over all the low multipoles one wishes to consider (e.g. the multipoles that have similar preferred axes) as in \citet{spaceoddity}: $\tilde{S}_{\rm tot}(\hat{\bf n}) =\sum_{\ell=2}^{\ell_{\rm max}} \tilde{S}_\ell (\hat{\bf n}).$ The parity estimator is redefined as $S(\hat{\bf n}) = \tilde{S}_{\rm tot}(\hat{\bf n}) - (\ell_{\rm max}-1)$, so that $\left<S\right> = 0$. The most even and odd mirror-parity directions for a given map can be considered by estimating \citep{spaceoddity}: 
\begin{eqnarray}S_+ = \frac{\max(S)-\mu(S)}{\sigma(S)}~{\rm and }~&S_- = \frac{|\min(S)-\mu(S)|}{\sigma(S)},\label{eq:sodd}\end{eqnarray}
where $\mu(S)$ and $\sigma(S)$ are the mean and standard deviation.

Others have also studied point parity with different statistics, e.g. \citet{Land:2005odd}, who did not find significant point parity in the first WMAP data release.  \citet{Kim:2010,Kim:2010b,Kim:2011}, however, find evidence of odd point parity in later WMAP renditions, and link this anomaly with the low level of correlations on the largest scales. 

\section{Validation of sparse inpainting to study large-scale anomalies}\label{sec:validation}
One can test for preferred axes directly on different renditions of WMAP data, for example, on ILC maps. However, these may be contaminated on large scales due to Galactic foregrounds \citep{ilc:w9}. Another approach is to use a different basis set than spherical harmonics, i.e. use a basis that is orthonormal on a cut sky \citep[e.g.][]{Rossmanith:2012}. Alternatively, one can use sparse inpainting techniques to reconstruct full-sky maps \citep{Plaszczynski:2012,Dupe:2011,Rassat:2012} or other inpainting methods, such as diffuse inpainting \citep{Zacchei:2011} or inpainting using constrained Gaussian realisations \citep{Bucher:2012,Kim:2012}. Any inpainting technique should be tested for potential biases, specifically for the masks and statistical tests one is interested in. Here we use the sparse inpainting techniques first described in \citet{Abrial2008} and \cite*{starck:book10} and refined in \cite*{Starck:2013} to reconstruct regions of missing data. The advantage of this method is that is does not assume the `true' map is either Gaussian or isotropic, yet it allows it to be \citep[see for e.g.][]{Starck:2013D}.

\citet{Rassat:2012} show sparse inpainting is a bias-free reconstruction method for the low quadrupole, quadrupole/octopole alignment, and octopole planarity tests. Here, we test whether sparse inpainting is a bias-free method for both the AoE and mirror parity. We calculate 1000 Gaussian random field realisations of WMAP7 best-fit cosmology using the WMAP7 temperature analysis mask ($f_{\rm sky}=0.78$).

\subsection{Recovering the mean interangle ($\theta$) with a realistic Galactic mask}\label{sec:aoe}
   
We compare the mean interangle $\theta$ for the statistic given in Equation \ref{eq:aoe} for $\ell=2-5$, for each map before and after inpainting, using $nside=128$ for the CMB maps. We find $\theta \sim 57 ^\circ\pm 9^\circ$ in our simulations before and after inpainting, i.e. what is expected in the case of isotropic axes and a Gaussian random field \citep{Land:2005ad}. After inpainting with the WMAP7 mask, we find that $(\theta_{\rm true} - \theta_{\rm inp}) \sim -0.55^\circ \pm 10.7^\circ$, showing there is no significant bias in the estimation of the mean interangle after sparse inpainting is applied. While the bias is small, the error bar on the mean interangle after sparse inpainting is not negligible ($10^\circ$). Following \cite*{Starck:2013}, we test the statistic on an optimistic \emph{Planck}-like mask with $f_{\rm sky}=0.93$ and find $(\theta_{\rm true}- \theta_{\rm inp}) \sim 0.17^\circ \pm 7.3^\circ$, showing we can expect better reconstructions with \emph{Planck} data and smaller masks.
	\vspace{-0.2cm}
\begin{table}[htbp]
   \centering

   \begin{tabular}{@{} lccc @{}} 
& True & After inp.&Bias\\
\hline 
$\theta$& $57.5^\circ \pm 9.2^\circ$&$57.0^\circ \pm 9.2^\circ$&$0.55^\circ \pm 10.7^\circ$\\
\hline
   	$S_+$&$2.59\pm 0.30$&$2.59\pm0.30$&$-0.00039\pm0.22$\\
	$S_-$&$2.81\pm0.35$&$2.82\pm0.35$&$-0.0049\pm0.26$\\
      \end{tabular}
   \caption{Mean interangle $\theta$ for the `Axis of Evil' statistic and even ($+$) and odd ($-$) mirror parity statistics $S_\pm$ before and after sparse inpainting on 1000 Gaussian random field realisations of CMB data and using the WMAP7 temperature analysis mask for the inpainted maps. The bias is taken by considering the difference $({\rm true} - {\rm inp}$).}
   \label{tab:parity_tests}
   	\vspace{-1.2cm}
\end{table}

\subsection{Recovering mirror parity statistics ($S\pm$) with a realistic Galactic mask}

To test for possible biases in the $S_\pm$ statistics after sparse inpainting, we calculate $S_+$ and $S_-$ for each CMB simulation before and after inpainting, setting $nside=8$ for the CMB maps, and $nside=64$ for the parity maps (calculated using Equation \ref{eq:paritymap}), as in \citet{spaceoddity}. As in Figure 6 of \citet{spaceoddity}, we find that the distributions of $S_+$ and $S_-$ populations do not change before and after inpainting (`True' and `After Inp.' in Table \ref{tab:parity_tests}).  We do not find any significant bias in the $S_\pm$ measurements (`Bias' column in Table \ref{tab:parity_tests}). 

Following \cite*{Starck:2013}, we also test the statistic on an optimistic \emph{Planck}-like mask with $f_{\rm sky}=0.93$ and find $\Delta S_+=-0.0021 \pm 0.081$ and $\Delta S_-=0.00091\pm 0.10$, showing we can expect significantly better reconstructions with future \emph{Planck} data and smaller masks.

\section{Recovering the primordial CMB}\label{app:isw}

\begin{table*}[htbp]
   \centering
   \begin{tabular}{@{} lccccccccc@{}} 
   \hline
   Map&Mean&$\ell=2$&&$\ell=3$&&$\ell=4$&&$\ell=5$&\\
   &interangle $\theta$ ($^\circ$) &$(b,l)$&$m$&$(b,l)$&$m$&$(b,l)$&$m$&$(b,l)$&$m$\\
      \hline
      \hline
      Before inpainting\\
      \hline
      1)\\
      TOH1&20.9&$(58.9, -103.4)$&$2$&$(61.9, -104.8)$&$3$&$(57.8, -164.0)$&$2$&$(47.7,-132.6)$&$3$\\
      W3 &51.9&$(-27.4, 3.3)$&$0$&$(62.3,-103.8)$&$3$&$(34.6,-132.2)$&$1$&$(47.4,-129.9)$&$3$\\
      W5&19.7&$(61.2,-121.7)$&$2$&$(62.3,-103.8)$&$3$&$(34.2,-131.8)$&$1$&$(47.4,-129.9)$&$3$\\
      W7 &20.4&$(62.7,-123.4)$&$2$&$(62.7,-104.0)$&$3$&$(33.9,-131.5)$&$1$&$(47.4, -130.7)$&$3$\\
      W9&19.1&$(60.1,-120.6)$&$2$&$(62.7,-105.2)$&$3$&$(34.2,-131.1)$&$1$&$(47,7,-130.2)$&$3$\\
          \hline
          2)\\
      TOH1 - kD&20.8&$(56.6, -106.5)$&$2$&-&-&-&-&-&-\\
      W3 - kD&20.3&$(62.7, -129.5)$&$2$&-&-&-&-&-&-\\
      W5- kD&18.9&$(57.8,-125.7)$&$2$&-&-&-&-&-&-\\
      W7 - kD&19.5&$(58.5,-127.6)$&$2$&-&-&-&-&-&-\\
        W9 - kD &18.5&$(57.0,-124.9)$&$2$&-&-&-&-&-&-\\
        \hline
      \hline
      After inpainting\\
      \hline
      3)\\
      TOH1 (inp) - kD&51.0&$(48.5,-116.4)$&$2$&$(29.3,82.6)$&$1$&$(57.8,-168.1)$&$2$&$(47.4,-140.0)$&$3$\\
      W3 (inp) - kD& 48.3&$(59.7,-140.0)$&$2$&$(31.7, 81.6)$&$1$&$(58.2,-165.9)$&$2$&$(47.0,-135.0)$&$3$\\
      W5 (inp) - kD&49.0&$(54.7,-135.0)$&$2$&$(31.4,81.9)$&$1$&$(58.2,-165.9)$&$2$&$(47.0,-135.0)$&$3$\\
      Dela W5 (inp) - kD&49.0&$(54.0,-140.6)$&$2$&$(28.0,82.6)$&$1$&$(58.9,-166.6)$&$2$&$(46.2,-135.8)$&$3$\\
      	W7 (inp) - kD&48.5&$(55.5,-138.9)$&$2$&$(30.7,82.6)$&$1$&$(58.2,-165.9)$&$2$&$(47.0,-135.8)$&$3$\\
	W9 (inp) - kD&16.3&$(54.7,-135.0)$&$2$&$(57.8,-116.4)$&$3$&$(58.2,-165.9)$&$2$&$(47.4,-135.4)$&$3$\\
             \hline
    4)\\
    TOH1 (inp) - kD - ISW &$56.1$&($36.4,-95.3$)&$2$&($19.5,18.6$)&$1$&($25.3,-19.3$)&$3$&($0.0,-36.2$)&$4$\\
W3 (inp) - kD - ISW  &$61.6$&($42.6,-91.8$)&$2$&($60.1,46.1$)&$2$&($25.0,-20.4$)&$3$&($0.0,-36.9$)&$4$\\
W5 (inp) - kD - ISW  &$53.8$&($43.0,-91.1$)&$2$&($23.0,-177.2$)&$2$&($24.6,-20.0$)&$3$&($0.3,-36.6$)&$4$\\
Dela W5 (inp) - kD - ISW &$53.7$&($44.6,-88.1$)&$2$&($23.6,-177.9$)&$2$&($25.0,-19.0$)&$3$&($0.3,-35.9$)&$4$\\
W7 (inp) - kD - ISW  &$55.8$&($42.6,-90.4$)&$2$&($19.2,18.3$)&$1$&($25.0,-20.4$)&$3$&($0.6,-36.2$)&$4$\\
W9 (inp) - kD - ISW  &$41.5$&($41.0,-90.4$)&$2$&($1.2,141.0$)&$1$&($25.0,-20.4$)&$3$&($0.6,-36.2$)&$4$\\
\hline
    

     \end{tabular}
   \caption{Preferred axes for multipoles $\ell=2-5$ for different WMAP CMB maps for $nside=128$.}
   \label{tab:aoe}
\end{table*}

Since statistical isotropy is predicted for the early Universe, analyses should focus on the primordial CMB, i.e. one from which secondary low-redshift cosmological signals have been removed. The observed temperature anisotropies in the CMB, $\delta_{\rm OBS}$, can be described as the sum of several components: 
\begin{equation} \delta_{\rm OBS} = \delta_{\rm prim} + \delta^{\rm total}_{\rm ISW} {\rm ~~~~on~large~scales},\end{equation}
where $\delta_{\rm prim}$ are the primordial temperature anisotropies, and $\delta^{\rm total}_{\rm ISW}$ the total secondary temperature anisotropies due to the late-time Integrated Sachs Wolfe (ISW) effect \citep[see Section 2.1 of][]{Rassat:2012}.

In practice, the temperature ISW field can be reconstructed in spherical harmonics, $\delta_{\ell m}^{\rm ISW}$, from the LSS field $g_{\ell m}$ \citep{Boughn:1998,Cabre:2007,Giannantonio:2008}: 
\begin{equation}
  \delta^{\rm ISW}_{\ell m} = \frac{C_{gT}(\ell)}{C_{gg}(\ell)} g_{\ell m},\label{eq:alm_isw}
\end{equation}
where $g_{\ell m}$ represent the spherical harmonic coefficients of a galaxy overdensity field $g(\theta, \phi)$.

The spectra $C_{gg}$ and $C_{gT}$ are the galaxy (g) and CMB (T) auto- and cross-correlations measured from the data \emph{or} their theoretical values \citep[see Section 2.2 of][]{Rassat:2012}.

\begin{table}[htbp]
   \centering
   \begin{tabular}{@{} lccc @{}} 
Map&$S_+$&$S_-$\\
\hline
1) Before inpainting \\
TOH1&3.25 (2.6\%)&3.15 (16\%)\\
WMAP3&2.88 (17\%)&2.81 (48\%)\\
WMAP5&2.93 (14\%)&2.85 (43\%)\\
WMAP7&2.93 (14\%)&2.88 (39\%)\\
WMAP9&3.00 (9.9\%)&2.93 (34\%)\\
\hline
2)After inpainting\\
TOH1 (inp) & 3.09 (5.7\%) &3.29 (10\%)\\
WMAP3 (inp)&3.31 (1.9\%)&3.39 (6.5\%)\\
WMAP5 (inp)&3.30 (2.0\%)&3.40 (6.1\%)\\
W5 Dela (inp)&3.41 (0.90\%)&3.57 (3.1\%)\\
WMAP7 (inp)& 3.34 (1.4\%)&3.55 (3.2\%)\\   
WMAP9 (inp) &3.20 (3.1\%)&3.54 (3.6\%)\\
\hline 
3)After inpainting\\ and ISW subtraction\\
TOH (inp)-ISW& 2.72 (34\%)  &  3.24 (12\%)   \\
 W3 (inp)-ISW&  2.81 (25\%)   &  3.21   (14\%) \\
 W5 (inp)-ISW&  2.78   (27\%)  & 3.23  (13\%)  \\
 W5 Dela (inp)-ISW&  2.77 (28\%) &   3.32 (10\%)\\   
 W7 (inp)-ISW&        2.85  (21\%)  &  3.32  (10\%)  \\
 W9 (inp) -ISW&        2.87  (20\%) &  3.28 (11\%)   \\

\end{tabular}
   \caption{Values of even ($S_+$) and odd ($S_-$) parity scores for $2<\ell<5$ for WMAP data from different years before (1) and after inpainting (2), and after subtraction of the ISW effect due to both 2MASS and NVSS galaxies (3). The occurrence for 1000 full-sky Gaussian random simulations is given in brackets. The kD quadrupole has been subtracted for all maps.}
   \label{tab:parity}
\end{table}

\end{document}